\documentclass[useAMS,usenatbib]{mn2e}
\usepackage{epsfig}
\usepackage{multirow}
\usepackage{enumerate}

\title[Correlations between ...]
{Correlations between the frequencies of twin kHz QPOs and spins of neutron stars in LMXBs}

\author[De-Hua Wang et al.]
{De-Hua Wang$^{1}\thanks{wangdh@gznu.edu.cn(DHW); zhangcm@bao.ac.cn(CMZ)}$, Cheng-Min Zhang$^{2,3,4\star}$, Jin-Lu Qu$^5$ and Yi-Yan Yang$^6$
\\
$^1$School of Physics and Electronic Science, Guizhou Normal University, Guiyang, 550001, China\\
$^2$National Astronomical Observatories, Chinese Academy of Sciences, Beijing, 100012, China\\
$^3$School of Physical Sciences, University of  Chinese Academy of Sciences, Beijing 100049, China\\
$^4$Key Laboratory of Radio astronomy, Chinese Academy of Sciences, Beijing 100012, China\\
$^5$Institute of High Energy Physics, Chinese Academy of Sciences, Beijing, 100049, China\\
$^6$School of Physics and Electronic Sciences, Guizhou Education University, Guiyang 550018, China\\
}
\date{Released 2002 Xxxxx XX}

\pagerange{\pageref{firstpage}--\pageref{lastpage}} \pubyear{2002}

\begin{document}

\maketitle

\label{firstpage}

\begin{abstract}

We investigate the correlation between the frequencies of the twin kilohertz quasi-periodic oscillations (kHz QPOs) and the neutron star (NS) spins in low mass X-ray binaries (LMXBs), based on the data sets  of  12 sources with simultaneously detected twin kHz QPOs and NS spins, and find that
the histogram of the ratio between the frequency difference of twin kHz QPOs ($\Delta\nu\equiv\nu_2-\nu_1$) and  NS spin
$\nu_{\rm s}$ shows a non-uniform distribution with a gap at  $\Delta\nu/\nu_{\rm s}\sim0.65$.
We try to classify the 12 sources into two categories according to this gap:
(I) The slow rotators with $\langle\nu_{\rm s}\rangle\sim311$\,Hz, XTE J1807.4-294, 4U 1915-05, IGR J17191-2821, 4U 1702-43, 4U 1728-34 and 4U 0614+09 follow a relation $\Delta\nu/\nu_{\rm s}>0.65$;
(II) The fast rotators with $\langle\nu_{\rm s}\rangle\sim 546$\,Hz, SAX J1808.4-3658, KS 1731-260, Aql X-1, 4U 1636-53, SAX J1750.8-2900 and 4U 1608-52 satisfy the relation $\Delta\nu/\nu_{\rm s}<0.65$.
However, the linear fittings of $\Delta\nu$ versus $\nu_{\rm s}$ relations of group (I) and (II) are unsatisfactory to ensure any  certain correlations.
We suggest that this phenomenon may arise from the fact that most measured kHz QPOs and spins satisfy the conditions of  $1.1\,\nu_{\rm s}\leq\nu_2<1300$\,Hz and $\Delta\nu$ decreasing with $\nu_2$. Apparently, the  diversified distribution of $\Delta\nu/\nu_{\rm s}$ refutes the simple  beat-frequency model, and the statistical correlations between  the twin kHz QPOs and NS spins may arise from  the magnetosphere-disk boundary environments, e.g., co-rotation radius and NS radius,  that modulate the occurrences of X-ray signals.  Furthermore, we also find that the distribution of the ratio of $\nu_2$ to $\nu_1$ clusters around the value of $\langle\nu_2/\nu_1\rangle\sim3:2$, which shows no obvious correlation with NS spins.

\end{abstract}


\section{Introduction}
\begin{table*}
\caption{The frequencies of the simultaneously detected twin kHz QPOs and NS spins.}
\label{QPOs}
\centering
\setlength{\tabcolsep}{5pt}
\begin{tabular}{@{}lcclccccl@{}}
\hline
\noalign{\smallskip}
Source (12)$^{a}$ & $\nu_1$$^{b}$ & $\nu_2$$^{c}$ & $\nu_{\rm s}$$^{d}$ & $\Delta\nu$$^{e}$ & $\langle\Delta\nu\rangle$$^{f}$ &  $\Delta\nu/\nu_{\rm s}$ & $\nu_2/\nu_1$ & References \\
 & (Hz) & (Hz) & (Hz) & (Hz) & (Hz) & & & \\
\noalign{\smallskip}
\hline
\noalign{\smallskip}
XTE J1807.4-294 & $106\sim370$ & $337\sim587$ & 191 (A) & $179\sim247$ & 197 & $0.94\sim1.29$& $1.51\sim3.18$  & [1, 13] \\ 
4U 1915-05 & $224\sim707$ & $514\sim1055$ & 270 (N) & $290\sim353$ & 299 & $1.07\sim1.31$ & $1.49\sim2.30$ & [2, 13] \\ IGR J17191-2821 & $681\sim870$ & $1037\sim1185$ & 294 (N) & $315\sim362$ & 349 & $1.07\sim1.23$ & $1.36\sim1.53$ & [3, 13] \\
4U 1702-43 & $722$ & $1055$ & 330 (N) & $333$ & 333 & $1.00$ & $1.46$ & [4, 13] \\
4U 1728-34 & $308\sim894$ & $582\sim1183$ & 363 (N) & $231\sim363$ & 341 & $0.64\sim1.00$ & $1.31\sim1.89$ & [5, 13] \\
SAX J1808.4-3658 & $435\sim567$ & $599\sim737$ & 401 (AN) & $164\sim195$ & 185 & $0.41\sim0.49$ & $1.30\sim1.39$ & [6, 13] \\
4U 0614+09 & $153\sim843$ & $449\sim1162$ & 415 (N) & $238\sim382$ & 317 & $0.57\sim0.92$ & $1.36\sim2.93$ & [7, 14] \\
KS 1731-260 & $898\sim903$ & $1159\sim1183$ & 524 (N) & $260\sim283$ & 272 & $0.50\sim0.54$ & $1.29\sim1.31$ & [8, 13] \\
Aql X-1 & $795\sim803$ & $1074\sim1083$ & 550 (AN) & $278\sim280$ & 279 & $\sim0.51$ & $\sim1.35$ & [9, 13] \\
4U 1636-53 & $529\sim979$ & $823\sim1228$ & 581 (N) & $230\sim341$ & 277 & $0.40\sim0.59$ & $1.23\sim1.56$ & [10, 13] \\
SAX J1750.8-2900 & $936$ & $1253$ & 601 (N) & $317$ & 317 & $0.53$ & $1.34$ & [11, 13] \\
4U 1608-52 & $473\sim867$ & $799\sim1104$ & 619 (N) & $225\sim326$ & 304 & $0.36\sim0.53$ & $1.26\sim1.69$ & [12, 13] \\
\noalign{\smallskip}
\hline
\noalign{\smallskip}
\end{tabular}
\begin{tabular}{@{}l@{}}
\begin{minipage}{160mm}
\begin{enumerate}[]
\item $^a$: The sources are listed in the order of NS spin frequency.
\item $^b$: $\nu_1$---Frequency of the lower kHz QPO.
\item $^c$: $\nu_2$---Frequency of the upper kHz QPO.
\item $^d$: $\nu_{\rm s}$---NS spin frequency inferred from periodic or nearly periodic X-ray
oscillations. A: accretion-powered millisecond pulsar. N: nuclear-powered millisecond pulsar.
\item $^e$: $\Delta\nu$---$\Delta\nu\equiv\nu_2-\nu_1$.
\item $^f$: $\langle\Delta\nu\rangle$---Weighted mean value of $\Delta\nu$ calculated by equation (\ref{mean_delt}).
\end{enumerate}
REFERENCES.---
[1] \citealt{Linares05}, \citealt{Zhang06b};
[2] \citealt{Boirin00};
[3] \citealt{Altamirano10a};
[4] \citealt{Markwardt99};
[5] \citealt{Di Salvo01}, \citealt{van Straaten02}, \citealt{Strohmayer96}, \citealt{Migliari03}, \citealt{Jonker00a}, \citealt{Mendez99};
[6] \citealt{van Straaten05},  \citealt{Wijnands03}, \citealt{Bult15};
[7] \citealt{van Straaten00}, \citealt{van Straaten02}, \citealt{Boutelier09};
[8] \citealt{Wijnands97};
[9] \citealt{Barret08};
[10] \citealt{Altamirano08b}, \citealt{Wijnands97a}, \citealt{Bhattacharyya10}, \citealt{Di Salvo03}, \citealt{Jonker00a}, \citealt{Jonker02a}, \citealt{Lin11}, \citealt{Sanna14};
[11] \citealt{Kaaret02};
[12] \citealt{van Straaten03}, \citealt{Barret05a}, \citealt{Jonker00a}, \citealt{Mendez98b};
[13] Reference in \citealt{Boutloukos08a};
[14] \citealt{Strohmayer08a}.
\end{minipage}
\end{tabular}
\end{table*}

The launch of the $Rossi\ X$-$Ray\ Timing\ Explorer$ ($RXTE$) has led to the discovery of Kilohertz quasi-periodic oscillations (kHz QPOs) in neutron star low mass X-ray binaries (NS-LMXBs) \citep{Strohmayer96,van der Klis96}. The frequencies of these QPOs cover the range from $\sim100$\,Hz to $\sim1300$\,Hz \citep{van der Klis00,van der Klis06} and correlate with other timing and spectral features \citep{Wijnands97a,Wijnands97b,Homan02,Kaaret98,Ford98c,Mendez99b, Ford00,Psaltis99,Belloni02}. These high-frequency QPOs usually appear in pairs (upper $\nu_2$ and lower $\nu_1$), with the frequencies show a nonlinear relation \citep{Belloni05,Zhang06a,Belloni07}, and so far there are over twenty LMXBs have shown twin kHz QPOs in accreting millisecond X-Ray pulsars (AMXP), atoll and Z sources \citep{Hasinger89,van der Klis00,van der Klis06,Wang14}.

It is believed that twin kHz QPOs may reflect the matter motion around the inner accretion disk dozens of kilometers away from the neutron star  (\citealt{Kluzniak90}, \citealt{van der Klis00} and references therein, \citealt{Zhang13,Wang15,Wang17}), which can be used to probe the strong gravitational field, the strong magnetic field around the neutron star, and constrain the NS $Mass-Radius$ relation \citep{Kluzniak90,van der Klis06,Wang13,Miller15}.
Although several theoretical interpretations are proposed, such as the relativistic precession \citep{Stella99a,Stella99b,Torok16}, the magnetohydrodynamic wave \citep{Zhang04} or non-linear resonance \citep{Kluzniak01,Abramowicz03a,Abramowicz03b} within an accretion disk, the sonic-point beat-frequency (SPBF) \citep{Miller98,Lamb01}, there is currently no consensus as to the origin of kHz QPOs.

The sonic-point beat-frequency model interprets the frequency separation of twin kHz QPOs to be  close to the NS spin \citep{Miller98,Lamb01}. However, the following observations show the averaged peak separations are found to be either close to the spin frequency or to its half \citep{Lamb01,Wijnands03,van der Klis06}. In addition, the resonance model also suggests the frequency of kHz QPOs to be related to the NS spin \citep{Lee04}. Thanks to the high timing revolution of $RXTE$, there are $\sim30$ NS-LMXBs  observed NS spin signals \citep{Burderi13,van der Klis16}, in which some sources have even been observed NS spin derivative (e.g. \citealt{Burderi06}) and twelve sources have also been detected the twin kHz QPOs, making it is possible to further analyze the correlation between the frequencies of twin kHz QPOs and NS spins. %
\begin{figure}
\centering
\includegraphics[width=8.4cm]{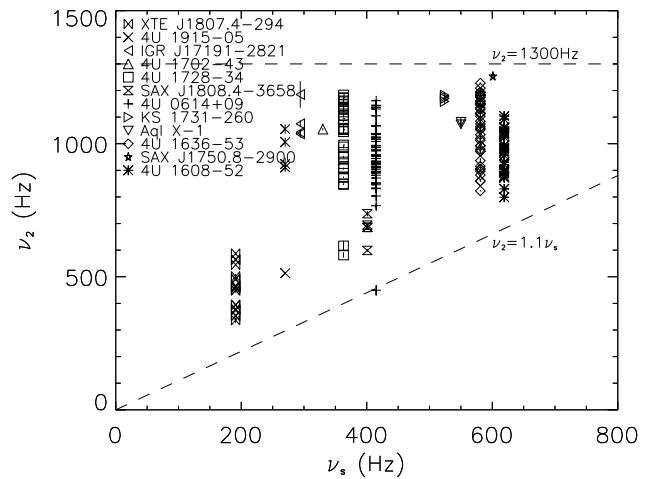}\\
\caption{Plot of the upper kHz QPO ($\nu_2$) versus NS spin ($\nu_{\rm s}$). The upper frequency $\nu_2$  of the 12 sources range at $\sim337-1253$\,Hz, which are constrained in the region of $\nu_{\rm s}\leq\nu_2<1300$\,Hz  by the dashed straight lines.}
\label{nu_u}
\end{figure}

\begin{figure*}
\centering
\includegraphics[width=8.4cm]{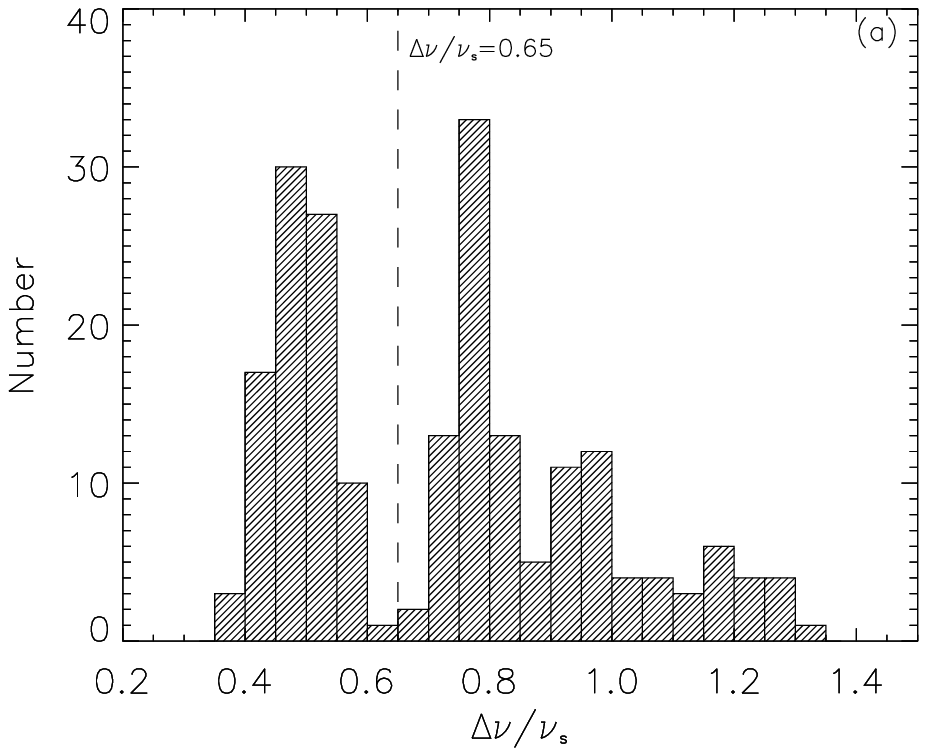}
\includegraphics[width=8.4cm]{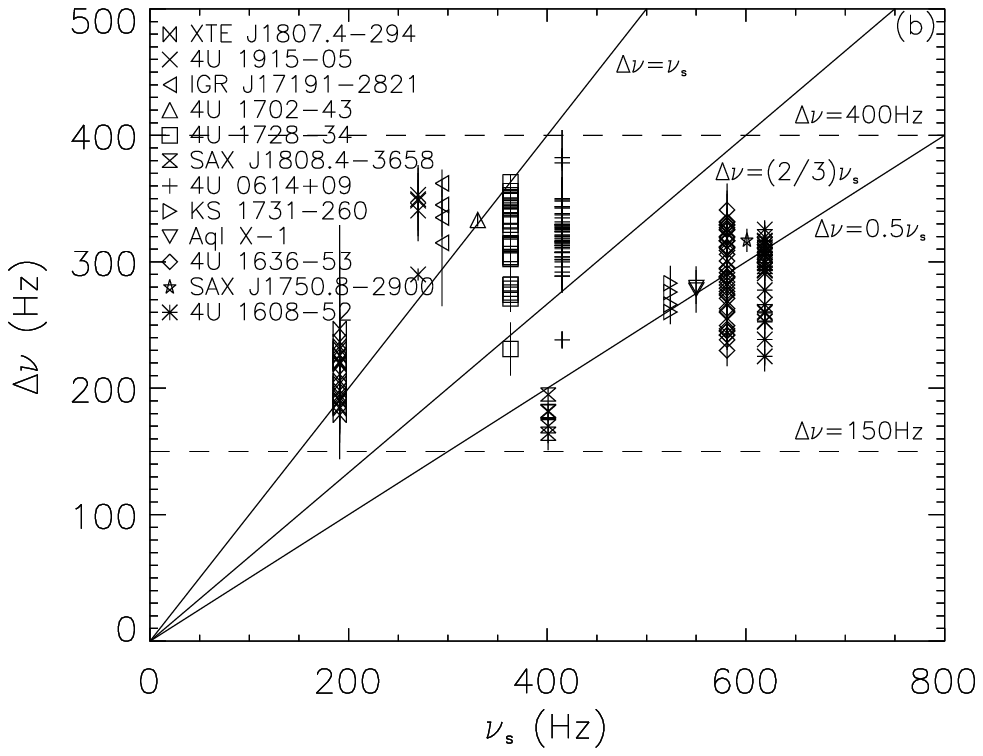}\\
\includegraphics[width=8.8cm]{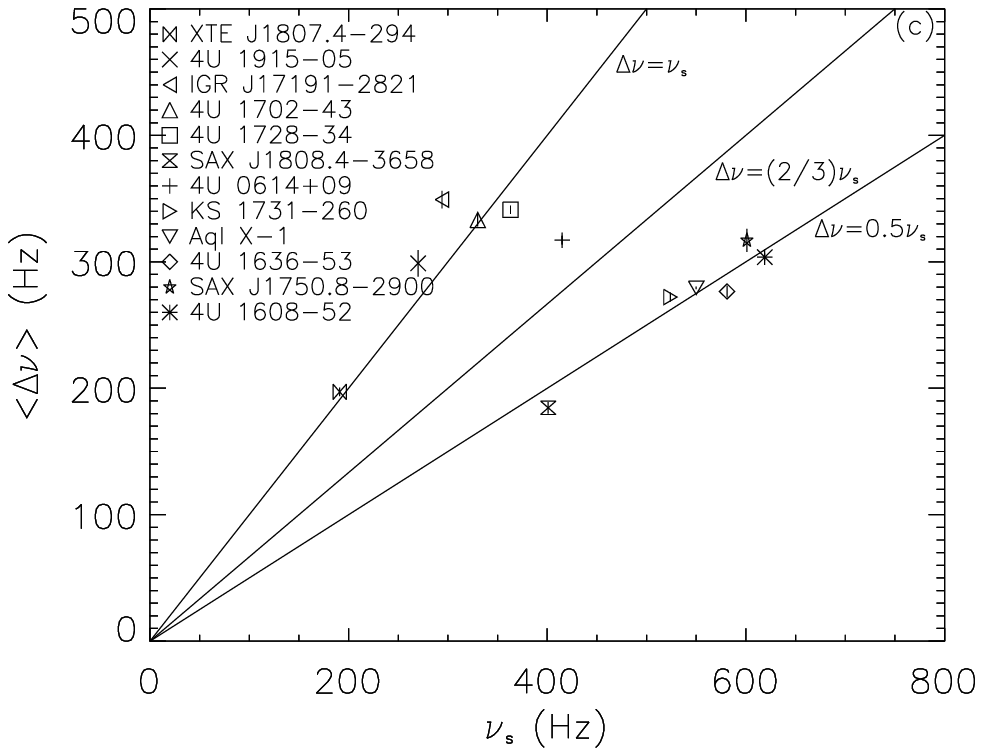}
\caption{
(a) The histogram of $\Delta\nu/\nu_{\rm s}$, where $\Delta\nu$ is the frequency difference between the upper and lower kHz QPOs ($\Delta\nu=\nu_2-\nu_1$)  and $\nu_{\rm s}$ is the NS spin.
(b) Plot of $\Delta\nu$ versus $\nu_{\rm s}$, where the lines of $\Delta\nu=0.5\,\nu_{\rm s}$, $\Delta\nu=(2/3)\,\nu_{\rm s}$ and $\Delta\nu=\nu_{\rm s}$ predicted by the resonance model \citep{Lee04} are also plotted.
(c) Plot of the weighted mean values of $\Delta\nu$ ($\langle\Delta\nu\rangle)$ versus $\nu_{\rm s}$.
}
\label{delt_nu}
\end{figure*}

The goal of this paper is to analyze the frequency correlation between the twin kHz QPOs and NS spins,  and its structure  is organized as follows: In $\S$ 2, we introduce the data of simultaneously detected twin kHz QPOs and NS spins  of 12 sources adopted in the  analysis, and their correlations are studied and  investigated in  $\S$ 3. Finally,  we present the discussions and conclusions in $\S$ 4.

\section{The published samples of the simultaneously detected twin kHz QPO   and NS spin frequencies}

We focus on the 12 NS-LMXBs which have been simultaneously detected the twin kHz QPOs and NS spins, and collect the data of these QPO and spin frequencies from the published literature. The samples contain 26 pairs of twin kHz QPOs from the two accreting millisecond X-ray pulsars (XTE J1807.4-294 and SAX J1808.4-3658), and the 177 ones from ten atoll sources. The NS spin frequencies of the 12 sources are inferred from either the periodic or nearly periodic X-ray burst oscillations \citep{Boutloukos08a,van der Klis06}.

The frequency range of the twin kHz QPOs from 12 NS-LMXBs are reported in Table \ref{QPOs}, from which it can be seen that the lower kHz QPOs show the frequency range of $\nu_1\sim106-979$\,Hz, while the upper kHz QPOs show the frequency range of $\nu_2\sim337-1253$\,Hz. Table \ref{QPOs} also shows the NS spin frequencies of the 12 sources, with the range of $\nu_{\rm s}\sim191-619$\,Hz and the average value of $\sim428$\,Hz. It is also noticed from Table\ref{QPOs} that XTE J1807.4-294 shows the relative lower frequencies of twin kHz QPOs and NS spin than other sources, i.e. $\nu_1\sim106-370$\,Hz, $\nu_2\sim337-587$\,Hz and $\nu_{\rm s}\sim191$\,Hz, respectively.

\begin{table}
\caption{Fitting results.}
\label{fitting}
\centering
\setlength{\tabcolsep}{3pt}
\begin{tabular}{@{}llll@{}}
\hline
\noalign{\smallskip}
Relation & Function & Parameter & $\chi^2/{\rm d.o.f.}$\\
\noalign{\smallskip}
\hline
\noalign{\smallskip}
\multirow{6}*{$\Delta\nu/\nu_{\rm s}$} & \multirow{2}*{$f(x)=$} & $a_1=32\pm11$ & \multirow{6}*{1.5} \\
 & & $b_1=0.78\pm0.02$ \\
 & \multirow{2}*{$a_1*{\rm e}^{-((x-b_1)/c_1)^2}+$} & $c_1=0.05\pm0.02$ \\
 & & $a_2=32\pm10$ \\
 & \multirow{2}*{$a_2*{\rm e}^{-((x-b_2)/c_2)^2}$} & $b_2=0.49\pm0.02$\\
 & & $c_2=0.08\pm0.03$ \\
\noalign{\smallskip}
\hline
\noalign{\smallskip}
\multirow{4}*{$\Delta\nu\sim\nu_{\rm s}$} & \multirow{2}*{(I) $\Delta\nu=a_1\,\nu_{\rm s}+b_1$} & $a_1=0.3\pm0.1$ & \multirow{2}*{5.3} \\
 & & $b_1=200\pm45$ & \\
 & \multirow{2}*{(II) $\Delta\nu=a_2\,\nu_{\rm s}+b_2$} & $a_2=0.5\pm0.1$ & \multirow{2}*{3.0} \\
 & & $b_2=-22\pm65$ & \\
\noalign{\smallskip}
\hline
\noalign{\smallskip}
\multirow{2}*{$\nu_2/\nu_1\sim\nu_{\rm s}$} & \multirow{2}*{$\nu_2/\nu_1=a\nu_{\rm s}^b$} & $a=2.4\pm0.6$ & \multirow{2}*{0.04} \\
 & & $b=-0.08\pm0.04$ \\
\noalign{\smallskip}
\hline
\end{tabular}
\end{table}
\section{Frequency correlation between the twin kHz QPOs and NS spins}

\subsection{The correlation between $\nu_2$ and $\nu_{\rm s}$}

\begin{figure}
\centering
\includegraphics[width=8.4cm]{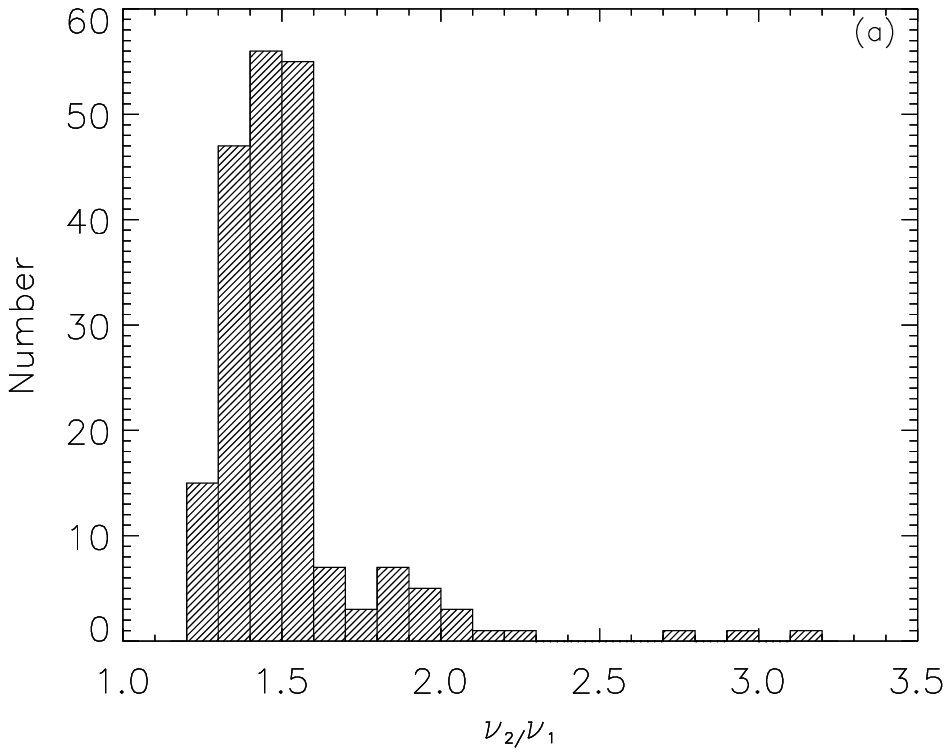}
\includegraphics[width=8.4cm]{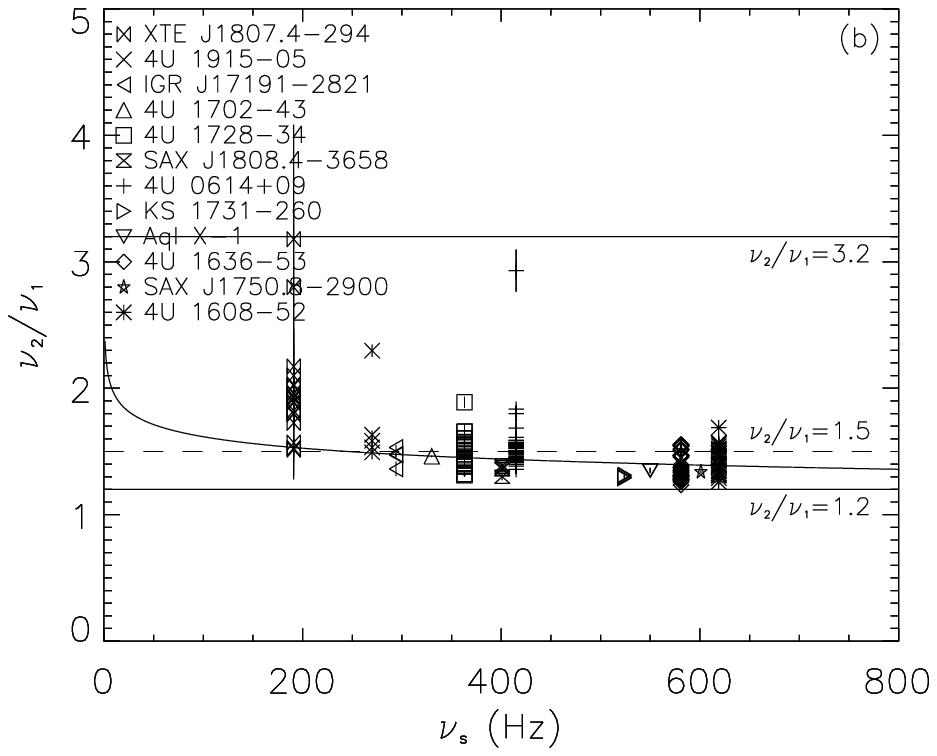}
\caption{
(a) The histogram of the frequency ratio of twin kHz QPOs ($\nu_2/\nu_1$).
(b) Plot of $\nu_2$/$\nu_1$ versus NS spin frequency ($\nu_{\rm s}$),
where the power-law curve fitting results are shown in Table \ref{fitting}.}
\label{rat_nu}
\end{figure}

Firstly, we try to probe the correlation between $\nu_2$ and $\nu_{\rm s}$ with the collected data of kHz QPOs and NS spins for  12 NS-LMXBs,
which is shown in Fig.\ref{nu_u}, where we notice that the upper limit of $\nu_2$  is around $\nu_2<1300$\,Hz.
It is also noticed from Table \ref{QPOs} and Fig.\ref{nu_u} that the NS spin frequency of each source is smaller than its upper kHz QPO frequencies, and the kHz QPO data are constrained in the region of $\nu_2\geq1.1\,\nu_{\rm s}$.
The distribution of the $\nu_2$ versus $\nu_{\rm s}$ relation is quite dispersive and there is no obvious concentrated phenomenon.

\subsection{The correlation between $\Delta\nu$ and $\nu_{\rm s}$}

The sonic-point beat-frequency model suggests the frequency difference between the upper and lower kHz QPOs ($\Delta\nu\equiv\nu_2-\nu_1$) may be related to the NS spin frequency $\nu_{\rm s}$ \citep{Miller98,Lamb01,Lamb03}.
The resonance model predicts that the ratio between $\Delta\nu$ and $\nu_{\rm s}$ ($\Delta\nu/\nu_{\rm s}$) is approximately equal to small integers, e.g. $\Delta\nu/\nu_{\rm s}\sim0.5$, $\sim2/3$ or $\sim1$, etc \citep{Lee04}.
However, there is currently no consensus as to these correlations. Here we calculate the values of $\Delta\nu$ for all the twin kHz QPOs and show the results in Table \ref{QPOs}. The range of $\Delta\nu$ is $\sim164-382$\,Hz, where SAX J1808.4-3658 and XTE J1807.4-294 show the lower $\Delta\nu$ values of $\sim164-195$\,Hz and $\sim179-247$\,Hz, respectively. We probe the correlation between $\Delta\nu$ and $\nu_{\rm s}$ from the following aspects:

We calculate the ratio between $\Delta\nu$ and $\nu_{\rm s}$ and show the results in Table \ref{QPOs}, where the range of $\Delta\nu/\nu_{\rm s}$ spans a lot of values from 0.36 to 1.31. Fig.\ref{delt_nu} (a) shows the histogram of $\Delta\nu/\nu_{\rm s}$, where the  bump distribution  is noticed. The Kolmogorov-Smirnov (K-S) test suggests that $\Delta\nu/\nu_{\rm s}$ distribution is not  uniform  at the 95\% confidence level,  implying that there may exist a  possible dependence of $\Delta\nu$ on $\nu_{\rm s}$.
It can be seen from Fig.\ref{delt_nu} (a) that there exists a distribution gap at  $\Delta\nu/\nu_{\rm s}\sim0.65$, based on which
we suspect that the 12 sources can be classified into two categories by this gap, and we also make a test of the double-gaussian function fitting that presents  the central values of the two peaks at  $\sim0.5$ and $\sim0.8$, respectively, as shown in Table \ref{fitting}. Furthermore, we find that  the $\Delta\nu/\nu_{\rm s}$ distribution from 0.7 to 1.3 is quite dispersive,  which is not sufficiently  obvious to classify them as a group.
Moreover, we also notice that the $\Delta\nu/\nu_{\rm s}$ distribution shown in Fig.\ref{delt_nu} (a) is partly similar to the expectation  of resonance model (see Fig.4 of \citet{Lee04}), which predicts a peak at $\Delta\nu/\nu_{\rm s}\sim0.5$ and other values,
 however its  predicted peak $\Delta\nu/\nu_{\rm s}\sim2/3$ is not found  in Fig.\ref{delt_nu} (a).

Fig.\ref{delt_nu} (b) shows the plot of $\Delta\nu$ versus $\nu_{\rm s}$, and a big  range of $\Delta\nu$ is found to be  $\sim150-400$\,Hz. So, the distribution of $\Delta\nu$ versus $\nu_{\rm s}$  is quite dispersive, however, the clustering phenomena separated by $\Delta\nu/\nu_{\rm s}\sim0.65$
can be seen, i.e. group (I) with $\Delta\nu/\nu_{\rm s}>0.65$: XTE J1807.4-294, 4U 1915-05, IGR J17191-2821, 4U 1702-43, 4U 1728-34 and 4U 0614+09; and group (II) with $\Delta\nu/\nu_{\rm s}<0.65$: SAX J1808.4-3658, KS 1731-260, Aql X-1, 4U 1636-53, SAX J1750.8-2900 and 4U 1608-52. Furthermore,
we try to find the non-biased linear correlations between $\Delta\nu$ and $\nu_{\rm s}$ of group (I) and (II),
 respectively,  by fitting them with the linear
functions. The fitting results are shown in Table \ref{fitting}, where the reduced $\chi^2$ of fitting on the two groups are
$\chi^2/{\rm d.o.f.}\sim5.3$ and $\chi^2/{\rm d.o.f.}\sim3.0$ respectively, implying the unsatisfactory fittings to show any  certain correlations.
In addition, Fig.\ref{delt_nu} (b) also shows the relations of $\Delta\nu=0.5\,\nu_{\rm s}$, $\Delta\nu=(2/3)\,\nu_{\rm s}$ and $\Delta\nu=\nu_{\rm s}$ by the straight lines, as  predicted by the resonance model \citep{Lee04}, however, it can be seen that the predicted correlations
seem to be partly consistent with the data, except  lacking  of data  around the line $\Delta\nu=(2/3)\,\nu_{\rm s}$.

For clarity, we calculate the weighted mean value of $\Delta\nu$, i.e. $\langle\Delta\nu\rangle$, of each source by the following equation:
\begin{equation}
\langle\Delta\nu\rangle=\frac{\sum\limits_{i=1}^N\Delta\nu_i/\sigma_i^2}{\sum\limits_{i=1}^N1/\sigma_i^2}
\label{mean_delt}
\end{equation}
where $\sigma_i$ is the error of $\Delta\nu_i$.
The $\langle\Delta\nu\rangle$ values are shown in Table \ref{QPOs}, and Fig.\ref{delt_nu} (c) shows the $\langle\Delta\nu\rangle$ versus $\nu_{\rm s}$ plot, where the clustering phenomena of the two category sources, i.e. group (I)---$\Delta\nu/\nu_{\rm s}\sim1$ and group (II)---$\Delta\nu/\nu_{\rm s}\sim0.5$, are more obvious, except for 4U 0614+09 whose $\langle\Delta\nu\rangle$ value lie near the lines $\langle\Delta\nu\rangle=0.76\,\nu_{\rm s}$.

\begin{figure}
\centering
\includegraphics[width=8.4cm]{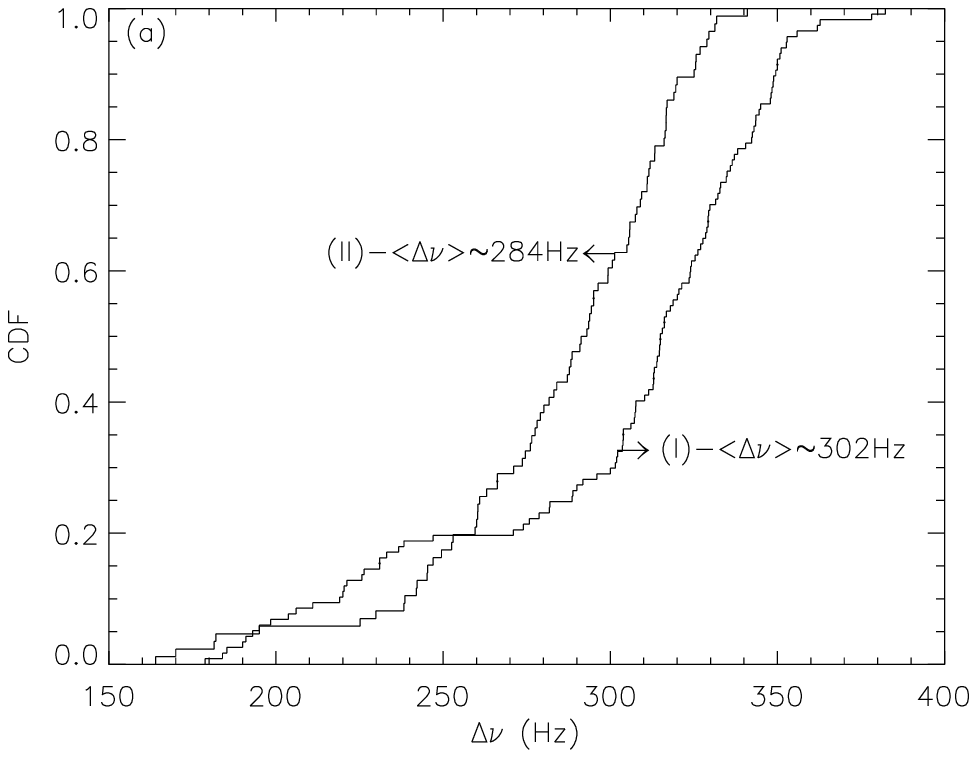}
\includegraphics[width=8.4cm]{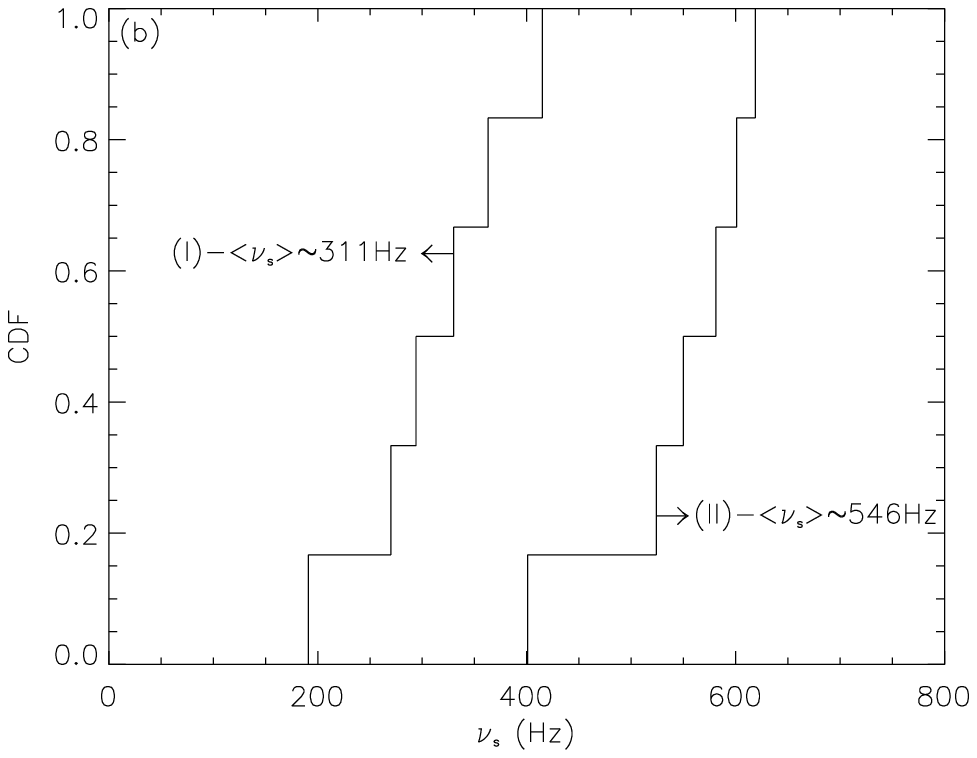}
\caption{
(a) The CDF curves of $\Delta\nu$ for group (I) and group (II) (see $\S$ 3.2).
(b) Similar to (a), but for the NS spins ($\nu_{\rm s})$.
}
\label{cdf_nu}
\end{figure}

In order to further investigate the difference between the two category sources, we compare their $\Delta\nu$ and $\nu_{\rm s}$ distributions, and show the corresponding cumulative distribution function (CDF) curves in Fig.\ref{cdf_nu} (a) and Fig.\ref{cdf_nu} (b), respectively. The K-S test shows that the $\Delta\nu$ data of the two category sources come from the different continuous distribution at the 95\% significance level, so does the $\nu_{\rm s}$ data. For group (I) and group (II), the mean values of $\Delta\nu$  are $\langle\Delta\nu\rangle\sim302$\,Hz and $\langle\Delta\nu\rangle\sim284$\,Hz, respectively, and the mean values of $\nu_{\rm s}$ are $\langle\nu_{\rm s}\rangle\sim311$\,Hz and $\langle\nu_{\rm s}\rangle\sim546$\,Hz, respectively.

\subsection{The correlation between $\nu_2$/$\nu_1$ and $\nu_{\rm s}$}

The non-linear resonance model by \citet{Abramowicz03a,Abramowicz03b,Abramowicz05} claimed that the
ratio relation between the frequencies of the pair high frequency QPOs in compact X-ray binaries to be approximately peaked about 3:2, which have been observed  in both stellar and intermediate black hole (BH) binaries \citep{Pasham14}, however this 3:2 phenomenon is not so sharp in the twin kHz QPOs of NS-LMXBs \citep{Belloni05,Belloni07}.

Here,  we calculated the values of $\nu_2/\nu_1$ and show the results in Table \ref{QPOs}, and find that the range of $\nu_2$/$\nu_1$ is $\sim1.23-3.18$ with the mean value of $\langle\nu_2$/$\nu_1\rangle\sim 3:2$. XTE J1807.4-294 shows the relative larger values of $\nu_2$/$\nu_1$ ($\sim1.51-3.18$) than those of other sources.
Fig.\ref{rat_nu} (a) shows the histogram of $\nu_2$/$\nu_1$, from which a ratio clustering phenomenon around 3:2 is noticed, as noted by the non-linear resonance model (see Fig.4 of \citealt{Abramowicz03a}).
Fig.\ref{rat_nu} (b) shows the diagram of  $\nu_2$/$\nu_1$ versus $\nu_{\rm s}$, where the distribution is quite dispersive. We fit the relation with a power-law function $\nu_2/\nu_1=a\nu_{\rm s}^b$ and show the results in Table \ref{fitting}. However, the reduced $\chi^2$ value of the fitting is quite poor ($\chi^2/{\rm d.o.f.}\sim0.04$),  which means that  there is no obvious correlation between $\nu_2/\nu_1$ and $\nu_{\rm s}$.

\section{Discussions and Conclusions}

Based on the data sets of the 12 sources with simultaneously detected twin kHz QPOs and NS spins, we investigate the correlation  between the twin kHz QPOs and the NS spins, and  find that there exists  a gap at $\Delta\nu/\nu_{\rm s}\sim0.65$ in the $\Delta\nu/\nu_{\rm s}$ distribution. The distribution of the ratio of $\nu_2$ to $\nu_1$ clusters around the mean value of $\langle\nu_2/\nu_1\rangle\sim3:2$, which shows no obvious correlation with NS spins.
The details of the conclusions are discussed and summarized as follows:

\begin{enumerate}[(1)]
\item The upper frequencies of twin kHz QPOs satisfy the conditions of $1.1\,\nu_{\rm s}\leq\nu_2<1300$\,Hz (see Table \ref{QPOs} and Fig.\ref{nu_u}), where the maximum value of  1300\,Hz may arise from the constrain by the NS stellar surface \citep{van der Klis06} and the relation $\nu_2\geq1.1\,\nu_{\rm s}$ may arise from the fact that the appearance of the twin kHz QPOs needs  a  critical orbital Keperlian  velocity of the accretion matter at the NS co-rotational radius \citep{Wang17}. It should also be noticed that the lower kHz QPO frequency $\nu_1$ are larger or smaller than the NS spin frequency $\nu_{\rm s}$ (see Table \ref{QPOs} and \citealt{Wang14}). As the NS cannot rotate faster than Keplerian frequency at the equator, so the relation of $\nu_2>\nu_{\rm s}$ and $\nu_1>~{\rm or}~<\nu_{\rm s} $ is compatible with the fact that the upper kHz QPO should be intimately involved in the Keperian orbital frequency, then the lower kHz QPO might not be directly given by the orbital frequency.

\item We focus on the statistical tests about how the relations proposed by various kHz QPO models compile with the detected  data. Firstly, the relativistic precession model predicts no correlation of the twin kHz QPO frequency difference with NS spin \citep{Stella99a,Stella99b,Torok16}.  The histogram of $\Delta\nu/\nu_s$ shows the bump phenomena (see Fig.\ref{delt_nu} (a)), which  is not compatible with a uniform distribution by the K-S test. So,
     we suspect that the frequencies of twin kHz QPOs and NS spins may obey an indirect relation, which may arise from the magnetosphere-disk   boundary environments, e.g., co-rotation radius and NS radius. Secondly, the sonic-point beat-frequency model interprets the frequency difference of twin kHz QPOs to be close to the NS spin frequency \citep{Miller98,Lamb01}. However, the $\Delta\nu/\nu_{\rm s}$ histogram in Fig.\ref{delt_nu} (a) shows a diversified distribution, which  obviously rejects the idea of claiming   $\Delta\nu/\nu_{\rm s}\sim1$, therefore, a  simple beat-frequency model should be refused. Thirdly, the forced resonance model by \citet{Lee04} predicts a peak at $\Delta\nu/\nu_{\rm s}\sim0.5$, $\sim2/3$ and $\sim1$, etc. It is  noticed that the $\Delta\nu/\nu_{\rm s}$ distributions of the detected data  at 0.5 and 1 seem to be consistent with the expectation of model,  however, the values around  2/3 are short of detected samples.

\item We try to classify the 12 sources into two categories based on the gap value at  $\Delta\nu/\nu_{\rm s}=0.65$: (I) As the slow rotators with $\langle\nu_{\rm s}\rangle\sim311$\,Hz, XTE J1807.4-294, 4U 1915-05, IGR J17191-2821, 4U 1702-43, 4U 1728-34 and 4U 0614+09 follow a relation $\Delta\nu/\nu_{\rm s}>0.65$; (II) As the fast rotators with $\langle\nu_{\rm s}\rangle\sim 546$\,Hz,  SAX J1808.4-3658, KS 1731-260, Aql X-1, 4U 1636-53, SAX J1750.8-2900 and 4U 1608-52 satisfy the relation $\Delta\nu/\nu_{\rm s}<0.65$ (see Fig.\ref{delt_nu} (b) and Fig.\ref{delt_nu} (c)). Because of  the  condition $\nu_2\geq1.1\,\nu_{\rm s}$  \citep{Wang17}  and  $\Delta\nu$ decreases  with $\nu_2$ \citep{van der Klis06,Zhang06a}, the slow rotators \citep{Lamb08} with the smaller $\nu_{\rm s}$ show the big  $\Delta\nu$, which causes the big  $\Delta\nu/\nu_{\rm s}$ value of $>0.65$. On the contrary, the fast rotators  with the big spin frequencies  $\nu_{\rm s}$ correspond to the small  $\Delta\nu$, which cause the smaller $\Delta\nu/\nu_{\rm s}$ value of $<0.65$.  However, it is not clear if the bimodal distribution  of $\Delta\nu/\nu_{\rm s}$ is possible, nor what physical process can interpret the gap around $\Delta\nu/\nu_{\rm s}\sim0.65$. If the correlation of $\Delta\nu$ versus $\nu_{\rm s}$ can be confirmed, it may be applied to estimate the NS spin frequencies in LMXBs.

\item As known, various bands of QPO frequencies in BH-LMXBs and NS-LMXBs follow the tight relations \citep{Belloni05,Belloni07},  and  a pair of  high frequency QPOs in stellar and intermediate black hole binaries follows, approximately,  a 3:2  ratio relation, which may be the particular phenomena from the innermost   stable circular orbit \citep{Abramowicz03a,Abramowicz03b,Pasham14}. In fact, there has been a suggestion in the literature for the 3:2 ratio as a parametric resonance between two particular modes of torus oscillations \citep{Bursa04,Kluzniak05,Kluzniak07}.
   However, we find that the distribution of the ratio of $\nu_2$ to $\nu_1$ of the 12 NS-LMXBs sources cluster around the value of $\sim3:2$ (see Fig.\ref{rat_nu} (a)), which shows no obvious correlation with NS spin.
   The statistical result of $\sim$3:2 QPO ratio relation shares the similar conclusion from the non-linear resonance model (see Fig.4 of \citealt{Abramowicz03a}),
 implying the $\sim$3:2  relation may be the common property of the compact X-ray binaries around some particular radius.
 The cause of why the ratios of the  pair  QPOs of  NS-LMXBs and BH-LMXBs  show  the $\sim$3:2 correlations is still unclear, and it  needs more efforts to uncover this secret  by analyzing more QPO data in the future detections.

\end{enumerate}

\section*{Acknowledgments}

This work is supported by the National Program on Key Research and Development Project (Grant No. 2016YFA0400803), the National Natural Science Foundation of China (Grant No. 11173034, No. 11673023 and No. 11703003), the Science and Technology Foundation of Guizhou Province (Grant No. J[2015]2113 and No. LH[2016]7226), the Doctoral Starting up Foundation of Guizhou Normal University 2014.
We are grateful for the critical comments and suggestions by the anonymous referee, which have significantly improved the quality of the paper.

\bsp

\label{lastpage}

\end{document}